\documentclass[journal,10pt]{IEEEtran}
\usepackage{geometry}
\usepackage{diagbox}
\usepackage{multirow}
\usepackage{booktabs}
\usepackage{graphicx}
\usepackage[fleqn]{amsmath}
\usepackage{cases}
\usepackage{subfigure}
\usepackage{stfloats}
\usepackage{epstopdf}
\usepackage{cite}
\usepackage{algorithm}
\usepackage{algorithmic}
\usepackage{tabu}

\usepackage{amssymb}
\usepackage{color}
\usepackage{xcolor}

\begin{document}
\title{Remote State Estimation with Posterior-Based Stochastic Event-Triggered Schedule}

\author{Zhongyao~Hu, Bo~Chen, Rusheng~Wang, Li~Yu
\thanks{Z. Hu, B. Chen, R. Wang and L. Yu are with Department of Automation, Zhejiang University of Technology, Hangzhou 310023, China. (email: bchen@aliyun.com).}
}
\markboth{Preprint}
{Shell \MakeLowercase{\textit{et al.}}: Bare Demo of IEEEtran.cls for Journals}
\maketitle

\begin{abstract}
This paper aims to study the state estimation problem under the stochastic event-triggered (SET) schedule. A posterior-based SET mechanism is proposed, which determines whether to transmit data by the effect of the measurement on the posterior estimate. Since this SET mechanism considers the whole posterior probability density function, it has better information screening capability and utilization than the existing SET mechanisms that only consider the first-order moment information of measurement and prior estimate. Then, based on the proposed SET mechanism, the corresponding exact minimum mean square error estimator is derived by Bayes rule. Moreover, the prediction error covariance of the estimator is proved to be bounded under moderate conditions. Meanwhile, the upper and lower bounds on the average communication rate are also analyzed. Finally, two different systems are employed to show the effectiveness and advantages of the proposed methods.
\end{abstract}


\vspace{-3pt}
\section{Introduction}
Remote state estimation has always been of interest in the long evolution of wireless applications from wireless sensor networks to cyber-physical systems \cite{W. Liu}. Under the remote state estimation scenarios, the energy consumption of sensors is mainly concentrated in remote transmission, and energy-limited batteries are likely to be the only power supply for the sensors. In this case, reducing sensor-to-estimator communication rate becomes quite important \cite{Det innovation 1}. Particularly, event-triggered schedule, which only transmits sensor data that meets the carefully designed criteria or importance metrics, provides a trade-off between estimation performance and energy consumption, and thus receives a lot of attention.

Generally, depending on whether the triggering threshold is a constant or a random variable, the event-triggered mechanisms can be divided into deterministic event-triggered (DET) mechanisms \cite{Det multiple point and set,Det innovation 1,Det innovation 2,Det innovation GSND,Det set,Det variance} and stochastic event-triggered (SET) mechanisms \cite{Sto SoD,Sto FIR,Sto initial,Sto attack,Sto energy harvesting,Sto information,Sto multi-sensor,Sto nonlinear,Sto packet drops,Sto smart sensor}. For the DET mechanisms, sensor data is transmitted to the estimator only if the triggering criterion goes beyond a certain threshold. This is a natural way to reduce the sensor-to-estimator transmission rate, because the sensor data that satisfies the triggering criterion tends to be more valuable. However, adopting the DET mechanisms will inevitably truncate the Gaussian probability density function (PDF) of the innovation sequence, in which case the exact minimum mean square error (MMSE) estimator is difficult to be obtained \cite{Det multiple point and set,Det innovation 1,Det innovation 2,Det set}. Although \cite{Det innovation GSND} designed the exact MMSE estimator under the innovation-based DET mechanism by utilizing the generalized closed skew normal distribution, its computational complexity grows with the passage of time, which makes this method difficult to be implemented in practice.
To overcome the drawback of the DET mechanisms, the SET mechanism was proposed in \cite{Sto initial}. In the SET mechanism, the triggering threshold is set to a random variable uniformly distributed over $[0,1]$, which subtly turns the transmission probability into a form similar to the Gaussian PDF and thus allows the Gaussianity of the innovation sequence to be preserved. Though the SET mechanism slightly increases the uncertainty of triggering, it brings great convenience to both theoretical analysis and practical implementation in return. With these benefits, the SET mechanism has received attention from many industries in just a few years after their proposals, such as multi-sensor data fusion \cite{Sto multi-sensor}, state estimation of networked systems \cite{Sto energy harvesting,Sto information,Sto packet drops,Sto smart sensor}, and secure estimation \cite{Sto attack}.

To pursue an exact and easy-to-implement MMSE estimator, this paper studies the state estimation problem under the SET mechanism. Note that while a lot of results have been developed on SET estimation, most of them are simply applications of existing SET mechanisms to different scenarios \cite{Sto attack,Sto energy harvesting,Sto information,Sto multi-sensor,Sto nonlinear,Sto packet drops,Sto smart sensor}. In contrast, few works have analyzed the SET mechanism itself, or have presented new ideas on the triggering criterion, which should be the most fundamental and important issue for the event-triggered schedule. Inspired by the work on continuous-time signal sampling \cite{SoD}, a send-on-delta(SoD)-based SET mechanism was introduced in \cite{Sto SoD}. However, this mechanism tends to perform poorly in non-smooth systems. The SET mechanism proposed in \cite{Sto initial} uses the size of innovation as a basis for triggering, and it outperforms the SoD-based SET mechanism as it is more in line with the update principle of Kalman filter (KF). Recently, a finite-impulse-response(FIR)-based SET mechanism was developed in \cite{Sto FIR}, which can essentially be understood as a trade-off between the SoD-based and innovation-based SET mechanisms. Unfortunately, while this method addresses the drawbacks of the SoD-based SET mechanism, it does not inherit the advantages of the innovation-based SET mechanism very well. In fact, even the best performing  innovation-based SET mechanism currently available still has two limitations: 1). It only considers the first-order moment information (i.e., mean) but ignores the second-order moment information (i.e., covariance) that reflects the estimation accuracy; 2). It considers the difference between the measurement and the priori estimate, but from a Bayesian perspective, the posterior information is the most central and comprehensive information in estimation, because it incorporates both the prior information and observation information. To overcome these two limitations, a new triggering idea will be presented in this paper. The main contributions of this paper are summarized as follows:
\begin{itemize}
\item[1)]
    It is proposed, for the first time, that whether an event is triggered should depend on the effect of the measurement on the posterior estimate. From a Bayesian point of view, posterior information is the core in estimation, and thus the proposed triggering mechanism can screen out those measurements that are important to the estimator better than the other SET mechanisms.
\item[2)]
    The exact MMSE estimator under the posterior-based SET mechanism is derived by means of Bayes rule, which has a similar recursive form to the standard KF and thus easy to implement in practice.
\item[3)]
    It is proved that the proposed estimator is asymptotically stable under moderate conditions. Moreover, expressions for the upper and lower bounds of the communication rate are derived, which can provide guidance for selecting parameter in the proposed SET mechanism.
\end{itemize}
\textbf{Notations:} ${\mathbb{R}}^r$ and ${\mathbb{R}}^{r\times s}$ denote the $r$ dimensional and $r\times s$ dimensional Euclidean spaces, respectively. ${\mathbb{N}}_{+}$ and ${\mathbb{N}}_{++}$ stand for nonnegative and positive integer number set, respectively. $\mathbb{E}[\cdot]$ denotes mathematical expectation. $p(\cdot)$ represents the PDF of a random variable, while $\mathrm{Pr}(\cdot)$ denotes the probability of a random event. $\mathrm{diag}\{\cdot\}$ stands for block diagonal matrix. $I$ stands for identity matrix. $\mathrm{Tr}(\cdot)$ and $\mathrm{Det}(\cdot)$ represent the trace and determinant of matrix, respectively. $\lambda_{max}(\cdot)$ and $\lambda_{min}(\cdot)$ represent the maximum and minimum eigenvalues respectively. For $X,Y\in{\mathbb{R}}^{r\times r}$, $X>Y$ and $X\geq Y$ mean that $X-Y$ is positive definite and positive semi-definite, respectively. $X^{1/2}$ is the square root of a matrix $X$, where $X\geq 0$. $\|x\|^2_{X}\triangleq x^TXx$ and $\|x\|^2\triangleq \|x\|^2_I$. $\phi(x;\overline{x},P)\triangleq\frac{1}{\sqrt{(2\pi)^{n}\mathrm{Det}(P)}}\exp\{-\frac{1}{2}(x-\bar{x})^TP^{-1}(x-\bar{x})\}$ represents Gaussian PDF, where $x\in{\mathbb{R}}^n$.
\section{Problem Formulation}
\subsection{Wasserstein distance}
Let $x$ and $y$ be two random variables with the same dimensions, the WD between them is expressed as \cite{WD}
\begin{equation}\begin{aligned}
&\mathcal{W}_{\Gamma}(x,y)
\triangleq (\inf\limits_{\gamma(x,y)}\iint||x-y||^2_{\Gamma}\gamma(x,y)dxdy)^{1/2}
\end{aligned}\end{equation}
where $\gamma(x,y)\in\{f(x,y)|\int f(x,y)dx=p(y),\int f(x,y)dy=p(x),f(x,y)\geq0\}$. The matrix $\Gamma>0$ does not exist in the initial WD and is introduced in this paper to adjust the weight of different state components and the communication rate. When $p(x)=\phi(x;\bar{x},P_x)$ and $p(y)=\phi(y;\bar{y},P_y)$, it can be shown that $\mathcal{W}_{\Gamma}(x,y)$ has analytic form \cite{WD Gaussian}
\begin{equation}\begin{aligned}
\mathcal{W}_{\Gamma}(x,y)=&(||\bar{x}-\bar{y}||^2_{\Gamma}\\
&+\mathrm{Tr}(P_x\Gamma+P_y\Gamma-2(SP_x\Gamma P_yS)^{1/2}))^{1/2},
\end{aligned}\end{equation}
where $S=\Gamma^{1/2}$.
\subsection{System Description}
Consider the general closed-loop event-triggered estimation system shown in Fig. 1 whose dynamic model is described by
\begin{equation}
\left\{ \begin{array}{l}
x_k=Ax_{k-1}+w_{k-1}\\
z_k=Cx_k+v_k
\end{array} \right.
\label {eq:1}
\end{equation}
where $x_k\in {\mathbb{R}}^n$ is the system state, $z_k\in \mathbb{R}^m$ is the measurement. $w_k$ and $v_k$ are uncorrected Gaussian white noises with covariance $Q>0$ and $R>0$, respectively. The initial state $x_0$ is Gaussian with mean $x_{0|0}$ and covariance $P_{0|0}>0$, and is uncorrelated with $w_k$ and $v_k$. $(C,A)$ is detectable. $\varsigma_k=1$ denotes the sensor transmits $z_k$ to the estimator and $\varsigma_k=0$ otherwise. In this case, at moment $k$, the available information set for the estimator is
\begin{equation}
\mathcal{I}_{1:k}=\{\mathcal{I}_1,\mathcal{I}_2,\cdots,\mathcal{I}_k\},
\end{equation}
where $\mathcal{I}_k\triangleq\{\varsigma_k,\varsigma_kz_k\}$ and $\mathcal{I}_{1:0}\triangleq\emptyset$. It is assumed that the feedback channel can reliably transmit data to the sensor \cite{Sto initial}.
   \begin{figure}[thpb]
      \centering
      \includegraphics[scale=0.35]{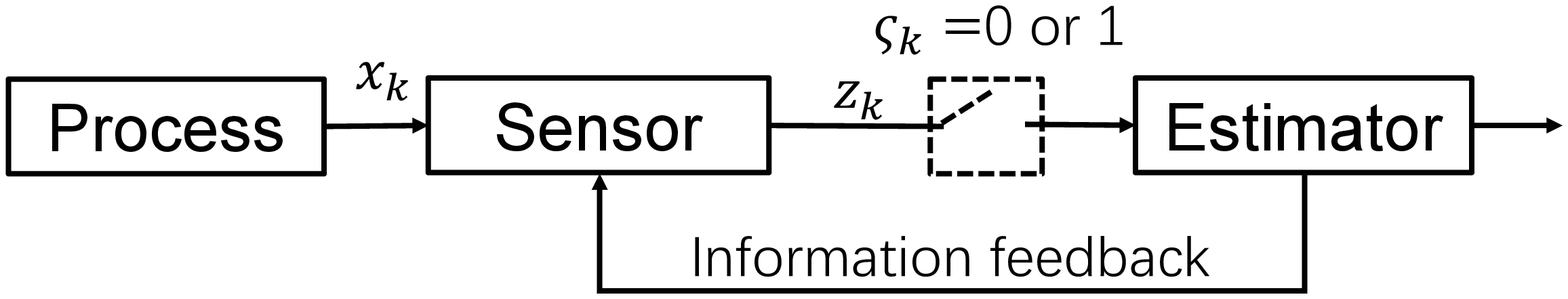}
      \caption{The closed-loop event-triggered estimation system. A typical example of such system in practice is remote state estimation based on IEEE 802.15.4/ZigBee protocol \cite{Sto initial, Zigbee}.}
      \label{Fig_1}
    \end{figure}

To facilitate later analysis, let us define
\begin{equation}
\left\{ \begin{array}{l}
x_{k|k-1}\triangleq \mathbb{E}[x_k|\mathcal{I}_{1:k-1}],\ x_{k|k}\triangleq \mathbb{E}[x_k|\mathcal{I}_{1:k}]\\
\tilde{x}_{k|k-1}\triangleq x_k-x_{k|k-1},\ \tilde{x}_{k|k}\triangleq x_k-x_{k|k}\\
P_{k|k-1}\triangleq \mathbb{E}[\tilde{x}_{k|k-1}\tilde{x}^T_{k|k-1}|\mathcal{I}_{1:k-1}]\\
P_{k|k}\triangleq \mathbb{E}[\tilde{x}_{k|k}\tilde{x}^T_{k|k}|\mathcal{I}_{1:k}],\ z_{k|k-1}\triangleq \mathbb{E}[z_k|\mathcal{I}_{1:k-1}]
\end{array}\right.
\end{equation}
where $x_{k|k-1}$ and $P_{k|k-1}$ are respectively mean and covariance of state prior PDF $p(x_k|\mathcal{I}_{1:k-1})$. $x_{k|k}$ and $P_{k|k}$ are respectively mean and covariance of state posterior PDF $p(x_k|\mathcal{I}_{1:k})$. $\tilde{x}_{k|k-1}$ and $\tilde{x}_{k|k}$ are respectively prior and posterior estimate error.

\textbf{Remark 1:} When the sensor transmits $z_k$ at each moment, the mean $x_{k|k}$ and error covariance $P_{k|k}$ of the posterior estimate are given by the standard KF \cite{Optimal filter}:
\begin{equation}\begin{aligned}
x_{k|k-1}=Ax_{k-1|k-1}
\end{aligned}\end{equation}
\begin{equation}\begin{aligned}
P_{k|k-1}=AP_{k-1|k-1}A^T+Q
\end{aligned}\end{equation}
\begin{equation}\begin{aligned}
K_k=P_{k|k-1}C^T(CP_{k|k-1}C^T+R)^{-1}
\end{aligned}\end{equation}
\begin{equation}\begin{aligned}
x_{k|k}=x_{k|k-1}+K_k(z_k-z_{k|k-1})
\end{aligned}\end{equation}
\begin{equation}\begin{aligned}
P_{k|k}=&(I-K_kC)P_{k|k-1}=(P_{k|k-1}^{-1}+C^TR^{-1}C)^{-1}
\end{aligned}\end{equation}
\subsection{Posterior-based SET Mechanism}
In the past, the innovation $z_k-z_{k|k-1}$ was often used as the basis for determining whether to trigger an event or not. Different from the innovation-based approach, a novel event-triggered idea will be proposed in this subsection. Let us start with a series of enlightening analyses. On one hand, from an information-theoretic point of view, the measurement $z_k$ is a function of the state $x_k$, and thus receiving $z_k$ implies more available information. On the other hand, it follows from (10) that the error covariance of KF is updated (i.e., $P_{k|k}<P_{k|k-1}$) whenever a measurement is received. From the above two aspects, it can be seen that regardless of the size of $z_k-z_{k|k-1}$, receiving a measurement is always better than not receiving one. This is an important feature of KF, but not considered in the innovation-based mechanism.

Let $x^1_k\triangleq x_k|\{\varsigma_k=1,z_k,\mathcal{I}_{1:k-1}\}$ and $x^0_k\triangleq x_k|\{\varsigma_k=0,\mathcal{I}_{1:k-1}\}$ stand for the posterior estimates when $z_k$ is transmitted and not transmitted, respectively. Then, it follows from the above analysis that, when the divergence between $x^0_k$ and $x^1_k$ is large, $z_k$ is more desired to be transmitted such that $x^0_k$ can be ``converted" to the better $x^1_k$. Conversely, when $x^0_k$ is very similar to $x^1_k$, the posterior estimates are almost the same whether $z_k$ is transmitted or not, in which case there is clearly no need to transmit $z_k$. This fact leads to the main idea of this paper, that is, \emph{the importance of $z_k$ depends on the divergence between $x^0_k$ and $x^1_k$}. Following this idea, we propose a novel posterior-based SET mechanism which is expressed as
\begin{equation}
\varsigma_k=
\left\{ \begin{array}{l}
0,\ \exp\{-\frac{1}{2}d(x^0_k,x^1_k)\}\geq\vartheta_k\\
1,\ \exp\{-\frac{1}{2}d(x^0_k,x^1_k)\}<\vartheta_k
\end{array} \right.
\end{equation}
where $\vartheta_k$ is uniformly distributed over $[0,1]$. Theoretically, $d(x^0_k,x^1_k)$ can be chosen to be any metric that measures the distance between random variables (e.g., $\mathcal{W}^2_{\Gamma}(x^0_k,x^1_k)$, $\mathbb{E}[\|x^1_k-x^0_k\|^2]$ and $\|\mathbb{E}[x^1_k]-\mathbb{E}[x^0_k]\|^2$). In this paper, we choose the WD as an example for the subsequent analysis.

Notice that, $\mathcal{W}_{\Gamma}(x^0_k,x^1_k)$ is a complex optimization problem and so far its analytic form exists only when both $x^0_k$ and $x^1_k$ are Gaussian. Unfortunately, $x^0_k$ and $x^1_k$ represent the outputs of the estimator and are unknown. In this case, the analytic form of $\mathcal{W}_{\Gamma}(x^0_k,x^1_k)$ cannot be obtained and the corresponding estimator is also impossible to be derived. To avoid this, we have to give a simplified and more ``analytic" metric
\begin{equation}\begin{aligned}
\mathcal{D}_{\Gamma}(x,y)\triangleq& (||\bar{x}-\bar{y}||^2_{\Gamma}\\
&+\mathrm{Tr}(P_x\Gamma+P_y\Gamma-2(SP_x\Gamma P_yS)^{1/2}))^{1/2}
\end{aligned}\end{equation}
where $\bar{x}\triangleq \mathbb{E}[x]$, $\bar{y}\triangleq \mathbb{E}[y]$, $P_x\triangleq \mathbb{E}[(x-\bar{x})(x-\bar{x})^T]$, and $P_y\triangleq \mathbb{E}[(y-\bar{y})(y-\bar{y})^T]$.
Although $\mathcal{W}_{\Gamma}(x,y)$ and $\mathcal{D}_{\Gamma}(x,y)$ seem to be different, it is fortunate that when both $x$ and $y$ are Gaussian, $\mathcal{W}_{\Gamma}(x,y)$ has the analytic form (2), which is the same as $\mathcal{D}_{\Gamma}(x,y)$.
Thus, if $x^0_k$ and $x^1_k$ can be shown to be both Gaussian, then $\mathcal{D}_{\Gamma}(x^0_k,x^1_k)$ will be equal to the Wasserstein distance $\mathcal{W}_{\Gamma}(x^0_k,x^1_k)$. Inspired by this, our subsequent attention is on proving $x^0_k$ and $x^1_k$ are Gaussian in the case $d(x^0_k,x^1_k)$=$\mathcal{D}^2_{\Gamma}(x^0_k,x^1_k)$.

Then, the problems to be solved in later sections are summarized as follows:
\begin{itemize}
\item Derive the analytic expressions of (11) and $p(x_k|\mathcal{I}_{1:k})$ under the case $d(x^0_k,x^1_k)=\mathcal{D}^2_\Gamma(x^0_k,x^1_k)$.
\item Analyze the performance of the proposed estimator to derive the stability conditions;
\item Analyze the average communication rate of the sensor to guide the selection of the adjustable parameter $\Gamma$.
\end{itemize}
\section{Main Results}
Before giving the main results of this paper, it is necessary to introduce the following lemmas. The proofs of Lemma 1 and Lemma 3 are presented in Appendix, the proof of Lemma 2 is straightforward and is therefore omitted.

\textbf{Lemma 1:} Let $E\geq 0\in {\mathbb{R}}^{r\times r}$ and $F>0\in {\mathbb{R}}^{r\times r}$, then one has $E-E(E+F)^{-1}E=F-F(E+F)^{-1}F\geq 0$.

\textbf{Lemma 2:} If the matrix sequence $\{E_k\}$ converges and $E_k\geq 0$, $\forall k\in \mathbb{N}_{++}$, then $\{E_k\}$ is bounded, i.e.,  there exists a matrix $\bar{E}$ such that $E_k\leq\bar{E}$, $\forall k\in {\mathbb{N}}_{++}$.

\textbf{Lemma 3:} When $E\geq H\geq 0$ and $F\geq D\geq  0$, one has $\mathrm{Det}(EF+I)\geq \mathrm{Det}(HD+I)$.
\subsection{MMSE Estimator Design}
In the existing works, estimators are derived with known schedulers. However, the posterior-based scheduler (11) is determined by the outputs $x^1_k$ and $x^0_k$ of the estimator, while the estimator is also determined by the posterior-based scheduler. This interconnection results in the explicit expressions for both the estimator and the scheduler are unknown, and thus the methodologies in \cite{Sto initial,Sto SoD,Sto FIR} cannot be directly applied here. A detailed analysis of the problem is given in Theorem 1.

\textbf{Theorem 1:} When $d(x^0_k,x^1_k)=\mathcal{D}^2_\Gamma(x^0_k,x^1_k)$, $p(x_k|\mathcal{I}_{1:k})$ is Gaussian with mean $x_{k|k}$ and covariance $P_{k|k}$ which are recursively calculated by the following Kalman-like form:
\begin{equation}\begin{aligned}
x_{k|k-1}=Ax_{k-1|k-1}
\end{aligned}\end{equation}
\begin{equation}\begin{aligned}
P_{k|k-1}=AP_{k-1|k-1}A^T+Q
\end{aligned}\end{equation}
\begin{equation}\begin{aligned}
K_k=P_{k|k-1}C^T(CP_{k|k-1}C^T+R)^{-1}
\end{aligned}\end{equation}
\begin{equation}\begin{aligned}
x_{k|k}=x_{k|k-1}+\varsigma_kK_k(z_{k}-Cx_{k|k-1})
\end{aligned}\end{equation}
\begin{equation}\begin{aligned}
P_{k|k}=&(I-K_kC)P_{k|k-1}+(1-\varsigma_k)\\
&\times K_k((CP_{k|k-1}C^T+R)^{-1}+K^T_k\Gamma K_k)^{-1}K^T_k.
\end{aligned}\end{equation}

\textbf{Proof:} Mathematical induction is used here to prove this theorem. First, the initial condition $p(x_0|\mathcal{I}_{1:0})=\phi(x_0;x_{0|0},P_{0|0})$ is Gaussian. Then, it is assumed that $p(x_{k-1}|\mathcal{I}_{1:k-1})=\phi(x_{k-1};x_{k-1|k-1},P_{k-1|k-1})$ is Gaussian. Under this case, it is obvious that the prior PDF is
\begin{equation}\begin{aligned}
p(x_{k}|\mathcal{I}_{1:k-1})=\phi(x_k;x_{k|k-1},P_{k|k-1}),
\end{aligned}\end{equation}
where the expressions for $x_{k|k-1}$ and $P_{k|k-1}$ are shown in (13) and (14).

It can be seen from (11) that $\varsigma_k$ is related to $x^0_k$ and $x^1_k$ only. Then, by recalling the definitions of $x^0_k$ and $x^1_k$, one knows that $\varsigma_k$ is essentially determined by $\mathcal{I}_{1:k-1}$ and $z_k$. Furthermore, since both $v_k$ and $w_k$ are white noises, the system (3) is a first-order hidden Markov model. Combining the above facts, one knows that $\varsigma_k$ is independent of $x_k$ given $(z_k,\mathcal{I}_{1:k-1})$. In this case, it follows from Bayes rule that
\begin{equation}\begin{aligned}
p(x^1_k)&=\frac{\mathrm{Pr}(\varsigma_k=1|x_k,z_k,\mathcal{I}_{1:k-1})p(x_k|z_k,\mathcal{I}_{1:k-1})}{\mathrm{Pr}(\varsigma_k=1|z_k,\mathcal{I}_{1:k-1})}\\
&=\frac{p(z_k|x_k)p(x_k|\mathcal{I}_{1:k-1})}{p(z_k|\mathcal{I}_{1:k-1})},
\end{aligned}\end{equation}
Then, similar to the standard KF, one has
\begin{equation}\begin{aligned}
p(x^1_k)=\phi(x_k;x^1_{k|k},P^1_{k|k})
\end{aligned}\end{equation}
where
\begin{equation}\begin{aligned}
x^1_{k|k}\triangleq x_{k|k-1}+K_k(z_k-Cx_{k|k-1})\nonumber
\end{aligned}\end{equation}
\begin{equation}\begin{aligned}
P^1_{k|k}\triangleq (I-K_kC)P_{k|k-1}\nonumber
\end{aligned}\end{equation}
where $K_k$ is given in (15).

When $\varsigma_k=0$, utilizing Bayes rule yields
\begin{equation}\begin{aligned}
p(x^0_k)&=p(x_k|\varsigma_k=0,\mathcal{I}_{1:k-1})\\
&=\frac{\mathrm{Pr}(\varsigma_k=0|x_k,\mathcal{I}_{1:k-1})p(x_k|\mathcal{I}_{1:k-1})}{\mathrm{Pr}(\varsigma_k=0|\mathcal{I}_{1:k-1})}.
\end{aligned}\end{equation}
It follows from the law of total probability that
\begin{equation}\begin{aligned}
&\mathrm{Pr}(\varsigma_k=0|x_k,\mathcal{I}_{1:k-1})\\
=&\int \mathrm{Pr}(\varsigma_k=0|v_k,x_k,\mathcal{I}_{1:k-1})\phi(v_k;0,R)dv_k.
\end{aligned}\end{equation}
According to (11) and (20), one has
\begin{equation}\begin{aligned}
&\mathrm{Pr}(\varsigma_k=0|v_k,x_k,\mathcal{I}_{1:k-1})\\
=&\alpha_k\exp\{-\frac{1}{2}(||K_kv_k-\xi_k)||^2_\Gamma\}
\end{aligned}\end{equation}
where
\begin{equation}\begin{aligned}
\left\{ \begin{array}{l}
\alpha_k=\exp\{-\frac{1}{2}\mathrm{Tr}(P^0_{k|k}\Gamma+P^1_{k|k}\Gamma-2(SP^0_{k|k}\Gamma P^1_{k|k}S)^{1/2})\}\\
\xi_k=x^0_{k|k}-x_{k|k-1}-K_k(Cx_k-Cx_{k|k-1})\\
x^0_{k|k}\triangleq \mathbb{E}[x_k|\mathcal{I}_{1:k-1},\varsigma_k=0]\\
P^0_{k|k}\triangleq \mathbb{E}[(x_k-x^0_{k|k})(x_k-x^0_{k|k})^T|\mathcal{I}_{1:k-1},\varsigma_k=0].
\end{array}\right.\nonumber
\end{aligned}\end{equation}
Notice that, $x^0_k$ is not directly related to $z_k$ and $\varsigma_k=0$ is constant. Thus, $x^0_{k|k}$ and $p^0_{k|k}$ remain invariant regardless of the value of $v_k$. In this case, the analytic form of (22) can be derived as
\begin{equation}\begin{aligned}
&\mathrm{Pr}(\varsigma_k=0|x_k,\mathcal{I}_{1:k-1})\\
=&\int \alpha_k\bar{\alpha}_k\hat{\alpha}_k
\exp\{-\frac{1}{2}||v_k-L^{-1}_kK^T_k\Gamma\xi_k||^2_{L_k}\}dv_k\\
=&\alpha_k\bar{\alpha}_k\hat{\alpha}_k(2\pi)^{n/2}\mathrm{Det}(L^{-1}_k)^{1/2}
\end{aligned}\end{equation}
where the last equation is obtained by the fact that the integral of Gaussian PDF is equal to 1, and
\begin{equation}\begin{aligned}
\left\{ \begin{array}{l}
L_k=K^T_k\Gamma K_k+R^{-1}\\
\bar{\alpha}_k=\frac{1}{(2\pi)^{n/2}\mathrm{Det}(R)^{1/2}}\\
\hat{\alpha}_k=\exp\{-\frac{1}{2}(\xi^T_k\Gamma\xi_k-\xi^T_k\Gamma K_kL^{-1}_kK^T_k\Gamma\xi_k)\}.
\end{array}\right.\nonumber
\end{aligned}\end{equation}

Then, by substituting (18) and (24) into (21) and taking a bit of tedious but straightforward algebra, one has
\begin{equation}\begin{aligned}
p(x^0_k)=\beta_k\bar{\beta}_k\exp\{-\frac{1}{2}(x_k-\eta_k)^TU_k(x_k-\eta_k)\}
\end{aligned}\end{equation}
where
\begin{equation}\begin{aligned}
\left\{ \begin{array}{l}
\beta_k=\frac{\alpha_k\bar{\alpha}_k}{\mathrm{Pr}(\varsigma_k=0|\mathcal{I}_{1:k-1})\mathrm{Det}(L_kP_{k|k-1})^{1/2}}\\
\bar{\beta}_k
=\exp\{\frac{1}{2}(\eta^T_kU_k\eta_k-\epsilon^T_k(\Gamma-\Gamma K_kL^{-1}_kK^T_k\Gamma)\epsilon_k\\
\ \ \ \ \ \ \ \ \ \ \ \ \ \ -x^T_{k|k-1}P^{-1}_{k|k-1}x_{k|k-1})\}\\
\epsilon_k=x^0_{k|k}-x_{k|k-1}+K_kCx_{k|k-1}\\
\eta_k=U^{-1}_k(V_k\epsilon_k+P^{-1}_{k|k-1}x_{k|k-1})\\
U_k=C^TK^T_k(\Gamma-\Gamma K_kL^{-1}_kK^T_k\Gamma)K_kC+P^{-1}_{k|k-1}\\
V_k=C^TK^T_k(\Gamma-\Gamma K_kL^{-1}_kK^T_k\Gamma).
\end{array}\right.\nonumber
\end{aligned}\end{equation}

Notice that both $\beta_k$ and $\bar{\beta}_k$ are constants with respect to $x_k$. Hence, according to the fact that Gaussian PDF integrals to 1, $\beta_k\bar{\beta}_k$ can be calculated by
\begin{equation}\begin{aligned}
\beta_k\bar{\beta}_k=\frac{1}{(2\pi)^{n/2}\mathrm{Det}(U^{-1}_k)^{1/2}}.
\end{aligned}\end{equation}
As a result, one has
\begin{equation}\begin{aligned}
&p(x^0_k)=\phi(x_k;\eta_k,U^{-1}_k)=\phi(x_k;x^0_{k|k},P^0_{k|k}).
\end{aligned}\end{equation}

Then, let $x^0_{k|k}=\eta_k$ and place the $x^0_{k|k}$-related terms on the left side of the equation, one can deduce that
\begin{equation}\begin{aligned}
U^{-1}_k(V_k-U_k)x^0_{k|k}=&U^{-1}_kV_k(x_{k|k-1}-K_kCx_{k|k-1})\\
&-U^{-1}_kP^{-1}_{k|k-1}x_{k|k-1}\\
(V_k-U_k)x^0_{k|k}=&(V_k-V_kK_kC-P^{-1}_{k|k-1})x_{k|k-1}\\
(V_k-U_k)x^0_{k|k}=&(V_k-U_k)x_{k|k-1},
\end{aligned}\end{equation}
which means $x^0_{k|k}=x_{k|k-1}$. Moreover, it follows from $(P^1_{k|k})^{-1}=P^{-1}_{k|k-1}+C^TR^{-1}C$ and Lemma 1 that $U_k$ is simplified to
\begin{equation}\begin{aligned}
U_k=(P^1_{k|k})^{-1}-C^TR^{-1}(K^T_k\Gamma K_k+R^{-1})^{-1}R^{-1}C.
\end{aligned}\end{equation}
Then, utilizing the matrix inverse lemma for $U^{-1}_k$ yields
\begin{equation}\begin{aligned}
&P^0_{k|k}=U^{-1}_k\\
=&P^1_{k|k}+K_{k}(K^T_k\Gamma K_k+(CP_{k|k-1}C^T+R)^{-1})^{-1}K^T_k.
\end{aligned}\end{equation}

Both $p(x^0_k)$ and $p(x^1_k)$ are Gaussian, and thus $p(x_k|\mathcal{I}_{1:k})$ is also Gaussian. Finally, summing the above results yields (13)-(17). This completes the proof. $\square$


Theorem 1 proves that $x^0_k$ and $x^1_k$ are Gaussian, and thus $\mathcal{D}_\Gamma(x^0_k,x^1_k)$ is equivalent to $\mathcal{W}_\Gamma(x^0_k,x^1_k)$. In this case, substituting the distributions of $x^0_k$ and $x^1_k$ into (11) gives
\begin{equation}
\varsigma_k=
\left\{ \begin{array}{l}
0,\ \exp\{-\frac{1}{2}(\|\varepsilon_k\|^2_{K^T_k\Gamma K_k}+\varrho_k)\}\geq\vartheta_k\\
1,\ \exp\{-\frac{1}{2}(\|\varepsilon_k\|^2_{K^T_k\Gamma K_k}+\varrho_k)\}<\vartheta_k
\end{array} \right.
\end{equation}
where $\varepsilon_k=z_k-z_{k|k-1}$ and $\varrho_k=\mathrm{Tr}(P^1_{k|k}\Gamma+P^0_{k|k}\Gamma-2(SP^0_{k|k}\Gamma P^1_{k|k}S)^{1/2})$.
As can be seen from (31), the posterior-based SET mechanism can be divided into two parts. $\|\varepsilon_k\|^2_{K^T_k\Gamma K_k}$ has a similar form to the innovation-based SET mechanisms, but differs in that $\|\varepsilon_k\|^2_{K^T_k\Gamma K_k}$ is reduced from $\|x^1_{k|k}-x^0_{k|k}\|^2_{\Gamma}$ and thus directly reflects the difference between the first order moments of posterior estimates. $\varrho_k$ represents the difference between the higher-order moments of posterior estimates, and is the most intuitive difference between the proposed SET mechanism (31) and previous SET mechanisms \cite{Sto initial,Sto SoD,Sto FIR,Det innovation GSND,Sto nonlinear,Sto multi-sensor,Sto information}. Therefore, the impact of $\varrho_k$ on (31) deserves to be analyzed (e.g., how much communication rate is induced by $\varrho_k$). The specific analysis is presented in Appendix C.



\textbf{Remark 2:}
Deriving $\phi(x_k;\eta_k,U^{-1}_k)$ inevitably makes use of information from the scheduler (11), which is determined by the unknown $x^0_k$. This means that $p(x^0_k)=\phi(x_k;\eta_k,U^{-1}_k)$ can actually be understood as an unsolved equation where $x^0_k$ is the unknown. In previous works \cite{Det innovation 1,Det innovation 2,Det set,Det multiple point and set,Det innovation GSND,Det variance,Sto initial,Sto SoD, Sto FIR,Sto nonlinear,Sto multi-sensor,Sto information}, the estimators were derived with the scheduler exactly known, so there is no need to solve such an equation.
\subsection{Performance Analysis}
For the sake of subsequent analysis, let us define
\begin{equation}\begin{aligned}
G_Y(X)\triangleq AXA^T+Q-AXC^T(CXC^T+Y)^{-1}CXA^T\nonumber
\end{aligned}\end{equation}
\begin{equation}\begin{aligned}
G^0_Y(X)\triangleq X,\ G^{k+1}_Y(X)\triangleq G_Y(G^k_Y(X))\nonumber
\end{aligned}\end{equation}
where $X>0$ and $Y>0$. Moreover, define
\begin{equation}\begin{aligned}
\left\{ \begin{array}{l}
\Theta_k\triangleq\frac{\lambda_{min}(\Gamma)}{\lambda_{max}(C^TC)}(I+\frac{1}{\delta_k}(C\underline{P}_kC^T)^{-1}(C\underline{P}_kC^T)^{-1}\\
\ \ \ \ \ \ \ +RR+\delta_k R(C\underline{P}_kC^T)^{-1}(C\underline{P}_kC^T)^{-1}R)^{-1}\\
\delta_k\triangleq\frac{\sqrt{Tr((C\underline{P}_kC^T)^{-1}(C\underline{P}_kC^T)^{-1})}}
{\sqrt{Tr(R(C\underline{P}_kC^T)^{-1}(C\underline{P}_kC^T)^{-1}R)}}
\end{array}\right.
\end{aligned}\end{equation}
\begin{equation}\begin{aligned}
\left\{ \begin{array}{l}
\Theta\triangleq\frac{\lambda_{min}(\Gamma)}{\lambda_{max}(C^TC)}(I+\frac{1}{\delta}(CP_lC^T)^{-1}(CP_lC^T)^{-1}\\
\ \ \ \ \ \ \ \ \ \ +RR+\delta R(CP_lC^T)^{-1}(CP_lC^T)^{-1}R)^{-1}\\
\delta\triangleq\frac{\sqrt{Tr((CP_lC^T)^{-1}(CP_lC^T)^{-1})}}
{\sqrt{Tr(R(CP_lC^T)^{-1}(CP_lC^T)^{-1}R)}}
\end{array}\right.
\end{aligned}\end{equation}

\textbf{Theorem 2:} Consider the system (3) with SET mechanism (31), if $C$ is full row rank, then $P_{k|k-1}$ in (14) is bounded by
\begin{equation}\begin{aligned}
0<\underline{P}_k \leq P_{k|k-1}\leq \bar{P}_k,\ \forall k\in {\mathbb{N}}_{++},
\end{aligned}\end{equation}
where
\begin{equation}\begin{aligned}
\underline{P}_k=G^{k-1}_R(AP_{0|0}A^T+Q),
\end{aligned}\end{equation}
\begin{equation}\begin{aligned}
\bar{P}_k=G_{R+\Theta^{-1}_k}(\bar{P}_{k-1}),
\end{aligned}\end{equation}
where $\bar{P}_1\triangleq AP_{0|0}A^T+Q$ and $\Theta_k$ is defined in (32). Particularly, the matrix sequences $\{\underline{P}_k\}$ and $\{\bar{P}_k\}$ are convergent, and their limits are $P_l$ and $P_u$, respectively. Here, $P_l$ and $P_u$ are respectively the unique positive definite solutions of
\begin{equation}\begin{aligned}
G_R(X)=X\ \mathrm{and}\ G_{R+\Theta^{-1}}(X)=X,
\end{aligned}\end{equation}
where $\Theta$ is defined in (33).

\textbf{Proof:} First, let us derive the lower bound for $P_{k|k-1}$. It is obvious that $P_{1|0}=AP_{0|0}A^T+Q\geq G^0_R(AP_{0|0}A^T+Q)$. Assume $P_{k|k-1}\geq G^{k-1}_R(AP_{0|0}A^T+Q)$. Then, one has
\begin{equation}\begin{aligned}
P_{k+1|k}=&AP_{k|k-1}A^T+Q-AP_{k|k-1}C^T(CP_{k|k-1}C^T\\
&+R)^{-1}CP_{k|k-1}A^T+(1-\varsigma_k)K_k\\
&\times ((CP_{k|k-1}C^T+R)^{-1}+K^T_k\Gamma K_k)^{-1}K^T_k\\
\geq& G_R(P_{k|k-1})\geq G^k_R(AP_{0|0}A^T+Q),
\end{aligned}\end{equation}
where the last inequality follows from the fact $G_Y(X_1)\geq G_Y(X_2)$ when $X_1\geq X_2\geq 0$. Moreover, notice that $G^1_R(0)\geq G^0_R(0)$, which further yields $G^2_R(0)=G^1_R(G^1_R(0))\geq G^1_R(0)$. By analogy, we can know that $\{G^k_R(0)\}$ is a monotonic non-decreasing matrix sequence, and $G^k_R(0)\geq G^1_R(0)=Q>0$ for all $k\in {\mathbb{N}}_{++}$. Then, by mathematical induction, it can be easily deduced that $G^k_R(AP_{0|0}A^T+Q)\geq G^k_R(0)>0$ for all $k\in {\mathbb{N}}_{++}$. Thus far, $0<\underline{P}_k \leq P_{k|k-1}$ has been proved.

Then, the upper bound of $P_{k|k-1}$ will be derived. Notice that $K^T_k\Gamma K_k$ is nonsingular when $C$ is full row rank, and thus we can rearrange $P_{k+1|k}$ in (14) by matrix inversion lemma as
\begin{equation}\begin{aligned}
P_{k+1|k}&=AP_{k|k-1}A^T+Q-AP_{k|k-1}C^T(CP_{k|k-1}C^T\\
&\ +R+(1-\varsigma_k)(K^T_k\Gamma K_k)^{-1})^{-1}CP_{k|k-1}A^T.
\end{aligned}\end{equation}
Moreover, it follows from Lemma 2.2 in \cite{Matrix inequality 1} and the expression of $K_k$ that
\begin{equation}\begin{aligned}
K^T_k\Gamma K_k\geq& \frac{\lambda_{min}(\Gamma)}{\lambda_{max}(C^TC)}(I+(CP_{k|k-1}C^T)^{-1}R)^{-1}\\
&\times(I+R(CP_{k|k-1}C^T)^{-1})^{-1}\geq \Theta_k,
\end{aligned}\end{equation}
where $\Theta_k$ is defined in (32). Then, to prove that $P_{k|k-1}\leq \bar{P}_k$, mathematical induction is used here. $P_{1|0}\leq \bar{P}_1$ is obvious. Suppose $P_{k|k-1}\leq \bar{P}_k$. Then, for $P_{k+1|k}$, one has
\begin{equation}\begin{aligned}
P_{k+1|k}=&G_{R+(1-\varsigma_k)(K^T_k\Gamma K_k)^{-1}}(P_{k|k-1})\\
\leq & G_{R+\Theta_k^{-1}}(P_{k|k-1})\leq G_{R+\Theta_k^{-1}}(\bar{P}_{k-1})=\bar{P}_{k+1},
\end{aligned}\end{equation}
which gives the upper bound.

Next, the convergence of the matrix sequences $\{\underline{P}_k\}$ and $\{\bar{P}_k\}$ will be proved. On one hand, it follows from the property of Riccati equation that $\{\underline{P}_k\}$ will convergent to the unique positive definite solution of $G_R(X)=X$ for all $\bar{P}_1\geq0$. On the other hand, it follows from the convergence of $\{\underline{P}_k\}$ that $\{\Theta^{-1}_k\}$ is also convergent and its limit $\lim_{k\rightarrow\infty}\Theta^{-1}_k=\Theta^{-1}$, where $\Theta$ is defined in (33). Then, according to Lemma 2, we know that there exists a matrix $\bar{\Theta}_1$ such that $\bar{\Theta}_1\geq\Theta^{-1}_k$ for all $k\in {\mathbb{N}}_{++}$. In this case, we can obtain
\begin{equation}\begin{aligned}
\bar{P}_k&\leq G_{R+\bar{\Theta}_1}(\bar{P}_{k-1})\leq \cdots \leq G^{k-1}_{R+\bar{\Theta}_1}(\bar{P}_1).
\end{aligned}\end{equation}
Obviously, $\{G^k_{R+\bar{\Theta}_1}(\bar{P}_1)\}$ is a convergent matrix sequence, and thus there exists a matrix $\bar{P}$ such that $\bar{P}_k\leq\bar{P}$ for all $k\in {\mathbb{N}}_{++}$. Moreover, it follows from (41) that $\bar{P}_k\geq Q>0$ for all $k\in {\mathbb{N}}_{++}$. Then, according to the definition of convergence, one can deduce that, for any $\varepsilon>0$, there is a $k_1\in {\mathbb{N}}_{++}$ such that
\begin{equation}\begin{aligned}
\Theta^{-1}-\varepsilon I<\Theta^{-1}_k<\Theta^{-1}+\varepsilon I,\ \forall k>k_1.
\end{aligned}\end{equation}
Then, let $k=k_1+k_2$, one has
\begin{equation}\begin{aligned}
G^{k_2}_{R+\Theta^{-1}-\varepsilon I}(Q)<\bar{P}_{k_1+k_2}&< G^{k_2}_{R+\Theta^{-1}+\varepsilon I}(\bar{P})
\end{aligned}\end{equation}
Notice that, when $k_1\rightarrow\infty$, $\varepsilon$ can tend to be 0. Therefore, let $k_1,k_2\rightarrow\infty$, one can deduce that
\begin{equation}\begin{aligned}
\lim_{k\rightarrow\infty} G^{k}_{R+\Theta^{-1}}(Q)\leq \lim_{k\rightarrow\infty}\bar{P}_k\leq \lim_{k\rightarrow\infty}G^{k}_{R+\Theta^{-1}}(\bar{P}).
\end{aligned}\end{equation}
According to the uniqueness of the solution of the discrete Riccati equation, one knows that $\bar{P}_k$ converges to $P_u$, which satisfies $G_{R+\Theta^{-1}}(P_u)=P_u$. This completes the proof. $\square$

\textbf{Theorem 3:} Consider the system (3) with SET mechanism (31), one has the following results:

1). The transformation probability $f_k\triangleq \mathbb{E}[\varsigma_k|\mathcal{I}_{1:k-1}]$ is
\begin{equation}\begin{aligned}
f_k=1-\frac{\exp\{-\frac{1}{2}\varrho_k\}}{\sqrt{\mathrm{Det}((CP_{k|k-1}C^T+R)K^T_k\Gamma K_k+I)}}.
\end{aligned}\end{equation}

2). When $C$ is row full rank, the average communication rate $\varsigma\triangleq\lim\limits_{k\rightarrow\infty}\mathbb{E}[\frac{\sum^{k}_{i=1}\varsigma_i}{k}]$ is bounded by $\underline{\varsigma}\leq\varsigma\leq\bar{\varsigma}$
where
\begin{equation}\begin{aligned}
\left\{ \begin{array}{l}
\underline{\varsigma}\triangleq 1-\frac{1}{\sqrt{\mathrm{Det}(\lambda_{min}(\Gamma) CP_lP_lC^T(CP_uC^T+R)^{-1}+I)}}\\
\bar{\varsigma}\triangleq 1-\frac{\exp\{-\mathrm{Tr}(P_u\Gamma)\}}{\sqrt{\mathrm{Det}(\lambda_{max}(\Gamma) CP_uP_uC^T(CP_lC^T+R)^{-1}+I)}}.
\end{array}\right.
\end{aligned}\end{equation}

\textbf{Proof:} 1). Utilizing the law of total probability yields
\begin{equation}\begin{aligned}
&\mathrm{Pr}(\varsigma_k=0|\mathcal{I}_{1:k-1})\\
=&\frac{\exp\{-\frac{1}{2}\varrho_k\}}{(2\pi)^{n/2}\sqrt{\mathrm{Det}(CP_{k|k-1}C^T+R)}}\\
&\times\int\exp\{-\frac{1}{2}\varepsilon^T_k(K^T_k\Gamma_kK_k\\
&\ \ \ \ \ \ \ \ \ \ +(CP_{k|k-1}C^T+R)^{-1})\varepsilon_k\}dz_k\\
=&\frac{\exp\{-\frac{1}{2}\varrho_k\}}{\sqrt{\mathrm{Det}((CP_{k|k-1}C^T+R)K^T_k\Gamma K_k+I)}}.
\end{aligned}\end{equation}
Then, it follows from $\mathbb{E}[\varsigma_k|\mathcal{I}_{1:k-1}]=1-\mathrm{Pr}(\varsigma_k=0|\mathcal{I}_{1:k-1})$ that (46) is obtained.

2).
Utilizing Corollary 3 in \cite{WD Gaussian} yields $\varrho_k\geq0$, which means
\begin{equation}\begin{aligned}
1\geq\exp\{-\frac{1}{2}\varrho_k\}.
\end{aligned}\end{equation}
Moreover, it follows from (34) and the definition of $\varrho_k$ that
\begin{equation}\begin{aligned}
\exp\{-\frac{1}{2}\varrho_k\}&\geq \exp\{-\frac{1}{2}\mathrm{Tr}((P^1_{k|k}+P^0_{k|k})\Gamma)\}\\
\geq & \exp\{-\mathrm{Tr}(P_{k|k-1}\Gamma)\}\geq \exp\{-\mathrm{Tr}(\bar{P}_k\Gamma)\}.
\end{aligned}\end{equation}

On the other hand, through simple simplifications we have
\begin{equation}\begin{aligned}
&(CP_{k|k-1}C^T+R)K^T_k\Gamma K_k\\
=&CP_{k|k-1}\Gamma P_{k|k-1}C^T(CP_{k|k-1}C^T+R)^{-1}.
\end{aligned}\end{equation}
Utilizing the conclusion of Theorem 2 yields
\begin{equation}\begin{aligned}
\lambda_{min}(\Gamma)C\underline{P}_k\underline{P}_kC^T&\leq CP_{k|k-1}\Gamma P_{k|k-1}C^T\\
&\ \ \leq \lambda_{max}(\Gamma)C\bar{P}_k\bar{P}_kC^T,
\end{aligned}\end{equation}
\begin{equation}\begin{aligned}
(C\bar{P}_kC^T+R)^{-1}&\leq(CP_{k|k-1}C^T+R)^{-1}\\
&\ \ \leq (C\underline{P}_kC^T+R)^{-1}.
\end{aligned}\end{equation}
Then, substituting (50)-(53) into (46) and utilizing Lemma 3 yields $\underline{f}_k\leq f_k\leq\bar{f}_k$, where
\begin{equation}\begin{aligned}
\left\{ \begin{array}{l}
\underline{f}_k\triangleq 1-\frac{1}{\sqrt{\mathrm{Det}(\lambda_{min}(\Gamma)C\underline{P}_k\underline{P}_kC^T(C\bar{P}_kC^T+R)^{-1}+I)}}\\
\bar{f}_k\triangleq1-\frac{\exp\{-\mathrm{Tr}(\bar{P}_k\Gamma)\}}{\sqrt{\mathrm{Det}(\lambda_{max}(\Gamma) C\bar{P}_k\bar{P}_kC^T(C\underline{P}_kC^T+R)^{-1}+I)}}.
\end{array}\right.
\end{aligned}\end{equation}
Notice that, $\mathbb{E}[\varsigma_k]$ satisfies
\begin{equation}\begin{aligned}
\mathbb{E}[\varsigma_k]=\int Pr(\varsigma_k=1|\mathcal{I}_{1:k-1})p(\mathcal{I}_{1:k-1})d\mathcal{I}_{1:k-1}.
\end{aligned}\end{equation}
In this case, substituting (55) into (57) yields $\underline{f}_k\leq \mathbb{E}[\varsigma_k]\leq\bar{f}_k$. Meanwhile, it follows from the convergence of matrix sequences $\{\underline{P}_k\}$ and $\{\bar{P}_k\}$ that the sequences $\{\underline{f}_k\}$ and $\{\bar{f}_k\}$ are also convergent, that is,
\begin{equation}\begin{aligned}
\lim\limits_{k\rightarrow\infty}\underline{f}_k=\underline{\varsigma},\ \lim\limits_{k\rightarrow\infty}\bar{f}_k=\bar{\varsigma},
\end{aligned}\end{equation}
where $\underline{\varsigma}$ and $\bar{\varsigma}$ are defined in (48). According to the definition of convergence, one knows that for any $\varepsilon>0$ there exists $k_3\in {\mathbb{N}}_{++}$ such that
\begin{equation}\begin{aligned}
\underline{f}_k>\underline{\varsigma}-\varepsilon,\ \bar{f}_k<\bar{\varsigma}+\varepsilon,\ \forall k>k_3.
\end{aligned}\end{equation}
Then, let $k=k_3k_4+k_3$, we can derive that
\begin{equation}\begin{aligned}
\frac{\sum^{k}_{i=1}\mathbb{E}[\varsigma_i]}{k}>
\frac{\sum^{k_3}_{i=1}\mathbb{E}[\varsigma_i]}{k}+\frac{k_3k_4(\underline{\varsigma}-\varepsilon)}{k},
\end{aligned}\end{equation}
\begin{equation}\begin{aligned}
\frac{\sum^{k}_{i=1}\mathbb{E}[\varsigma_i]}{k}<
\frac{\sum^{k_3}_{i=1}\mathbb{E}[\varsigma_i]}{k}+\frac{k_3k_4(\bar{\varsigma}+\varepsilon)}{k}.
\end{aligned}\end{equation}
Finally, let $k_3,k_4\rightarrow\infty$ and $\varepsilon\rightarrow 0$, (47) is obtained.
This completes the proof. $\square$

\section{Simulation Examples}
\subsection{Target Tracking Systems}
Consider a classical target tracking system whose state space-model is given by \cite{Sto initial}
\begin{equation}\begin{aligned}
x_k=\begin{bmatrix}
{1} & {T} & {T^2}\\
{0} & {1} & {T}\\
{0} & {0} & {1}
\end{bmatrix}x_{k-1}+w_{k-1},\ z_k=\begin{bmatrix}
{1} & {0} & {0}\\
{0} & {0} & {1}
\end{bmatrix}x_k+v_k
\nonumber
\end{aligned}\end{equation}
where $x_k=[s_k,\dot{s}_k,\ddot{s}_k]$. $s_k$, $\dot{s}_k$ and $\ddot{s}_k$ are respectively the position, velocity and acceleration of the target. $T=0.25s$ is the sampling period. $w_k$ and $v_k$ are white Gaussian noises, and their covariances are respectively
\begin{equation}\begin{aligned}
&Q=2a\sigma_m^2\begin{bmatrix}
{T^5/20} & {T^4/8} & {T^3/6}\\
{T^4/8} & {T^3/3} & {T^2/2}\\
{T^3/6} & {T^2/2} & {T}
\end{bmatrix}
,\
R=\begin{bmatrix}
{\sigma^2_p} & {0}\\
{0} & {\sigma^2_a}
\end{bmatrix},
\nonumber
\end{aligned}\end{equation}
where $a=5$, $\sigma_m=\sigma_a=0.1$, $\sigma_p=1$. The parameter matrix $\Gamma$ is set as $c\times\mathrm{diag}(4,1,1)$ where $c$ is an adjustable scalar.


When the parameter $c$ takes different values, the upper bound $\bar{\varsigma}$ and lower bound $\underline{\varsigma}$ of the average communication rate $\varsigma$ are shown in Fig. 2, which can provide guidance for selecting parameter $c$. Specifically, if the desired transformation rate is $r$, then $c$ will necessarily be taken to be on the interval $[\bar{c}(r),\underline{c}(r)]$, where $\bar{c}(r)$ and $\underline{c}(r)$ represent the values of $c$ that make $\bar{\varsigma}$ and $\underline{\varsigma}$ equal to $r$, respectively. For example, if the desired communication rate $\varsigma$ is $0.6$, then $c$ should fall on the interval $[1.6,76]$, as shown in the enlarged box in Fig. 2.
\begin{figure}[thpb]
      \centering
      \includegraphics[scale=0.6]{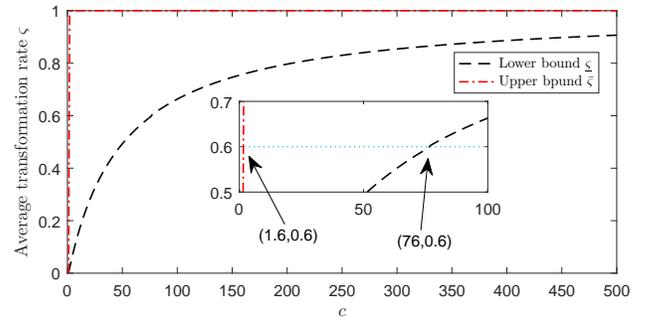}
      \caption{Upper and lower bounds of the average communication rate.}
      \label{O2_content}
\end{figure}


To demonstrate the advantages of the proposed posterior-based SET KF (PSET-KF), we compare it with the innovation-based SET KF (ISET-KF) in \cite{Sto initial}, SoD-based SET KF (SSET-KF) in \cite{Sto SoD}, FIR-based SET KF (FSET-KF) in \cite{Sto FIR}, and variance-based KF (V-KF) in \cite{Det variance}. Moreover, the parameter matrices in the ISET-KF, SSET-KF and FSET-KF are similarly chosen to be $c\times\mathrm{diag}(4,1)$. Let us define
\begin{equation}\begin{aligned}
\left\{ \begin{array}{l}
E(k)\triangleq\frac{1}{k}\sum^{k}_{j=1}\sqrt{\sum^M_{i=1}\frac{\|x_j(i)-\hat{x}_j(i)\|^2}{M}}\\
T(k)\triangleq\frac{1}{k}\sum^{k}_{j=1}\sum^M_{i=1}\frac{P_{j|j}(i)}{M}.
\end{array}\right.\nonumber
\end{aligned}\end{equation}
to evaluate the algorithms' estimation performance where $x_j(i)$ and $\hat{x}_j(i)$ are respectively the true and estimated states during the $i$th Monte Carlo experiment, $M=200$ is the total number of Monte Carlo experiments.
Then, Fig. 3 shows the estimation performances of the above methods for different communication rates, and the values of adjustable parameters are presented in Table I. From Fig. 3, it can be seen that
the proposed PSET-KF always outperforms other algorithms under different communication rates. Notice that, both PSET-KF, ISET-KF, SSET-KF, FSET-KF, and V-KF are optimal MMSE estimators, only the triggering ideas adopted are different. Therefore, the better estimation performance of PSET-KF essentially benefits from the fact that the posterior-based SET mechanism can better screen out those valuable measurements.
\begin{table*}
\centering
\caption{The relationship between the adjustable parameters and the communication rates}
\begin{tabular}{|c|c|c|c|c|c|c|c|c|c|c|}
\hline
                      & \diagbox{adjustable parameter}{communication rate} & 0.1  & 0.2  & 0.3  & 0.4  & 0.5  & 0.6  & 0.7  & 0.8  & 0.9    \\
\hline
\multirow{6}{*}{IV-A} & PSET-KF                                 & 0.06 & 0.62 & 2.3  & 5.9  & 12   & 24   & 45   & 88   & 220    \\
\cline{2-11}
                      & ISET-KF ($\times 10^{-1}$)               & 0.25 & 0.89 & 1.9  & 3.5  & 6    & 10.5 & 20   & 44   & 140    \\
\cline{2-11}
                      & SSET-KF ($\times 10^{-4})$               & 0.02 & 0.18 & 0.6  & 1.5  & 3.3  & 6    & 15   & 38   & 190    \\
\cline{2-11}
                      & FSET-KF ($\times 10^{-1}$)               & 0.06 & 0.25 & 0.6  & 1.2  & 2.2  & 3.8  & 7.5  & 16.5 & 52.9   \\
\cline{2-11}
                      & \multirow{2}{*}{V-KF ($\times 10^{-3}$)} & 2700 & 690  & 330  & 200  & 113  & 70   & 35   & 21   & 0.01   \\
\cline{3-11}
                      &                                         & 300  & 150  & 55   & 30   & 26   & 20   & 20   & 0.8  & 0.001  \\
\hline
\multirow{6}{*}{IV-B} & PSET-KF ($\times 10^{3}$)                & 0.72  & 3   & 7   & 14.5  & 25  & 42  & 69  & 130 & 300   \\
\cline{2-11}
                      & ISET-KF                                 & 0.9  & 2.1  & 3.8  & 6.3  & 10.5 & 17   & 29   & 55   & 150    \\
\cline{2-11}
                      & SSET-KF                                 & 0.4  & 1.1  & 1.9  & 3.1  & 5.2  & 8.8  & 15   & 30   & 80     \\
\cline{2-11}
                      & FSET-KF                                 & 0.15 & 0.5  & 1.2  & 2.2  & 4    & 6.8  & 12   & 24   & 60     \\
\cline{2-11}
                      & \multirow{2}{*}{V-KF ($\times 10^{-3}$)} & 5.5  & 2.9  & 1.8  & 1.44 & 0.96 & 0.68 & 0.44 & 0.28 & 0.14   \\
\cline{3-11}
                      &                                         & 0.89 & 0.45 & 0.27 & 0.22 & 0.14 & 0.1  & 0.06 & 0.04 & 0.02   \\
\hline
\end{tabular}
\end{table*}
\begin{figure}[thpb]
      \centering
      \includegraphics[scale=0.7]{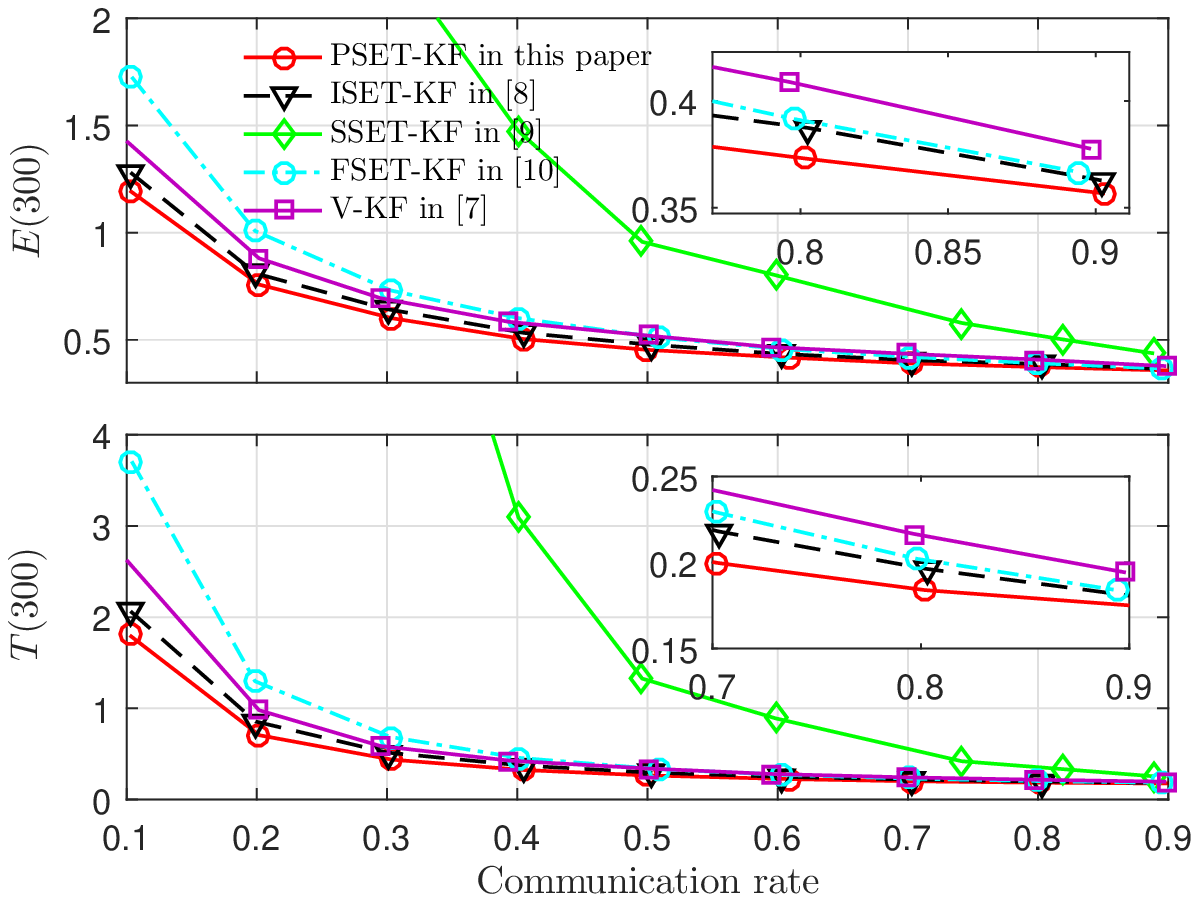}
      \caption{The estimation performances $E(300)$ and $T(300)$ of PSET-KF, ISET-KF, SSET-KF, FSET-KF and V-KF versus communication rate.}
\end{figure}

\subsection{Spring-Mass Systems}
To demonstrate the advantages of the proposed method more comprehensively, in this subsection we consider a spring-mass system (a typical oscillating system), whose differential equations can be written as
\begin{equation}\begin{aligned}
m_1\ddot{x}_1(t)=&-k_1(x_1(t)-x_2(t))+w_1(t)\\
m_2\ddot{x}_2(t)=&k_1(x_1(t)-x_2(t))-k_2x_2(t)+w_2(t)
\nonumber
\end{aligned}\end{equation}
where $x_i(t)$, $\dot{x}_i(t)$ and $\ddot{x}_i(t)$ represent the displacement, velocity and acceleration of the $i$th ($i=1,2$) object respectively. $m_i$ represents the mass of $i$-th object, $k_i$ is the $i$th spring factor. $w_1(t)$ and $w_2(t)$ are process noises with covariance $\sigma_1^2$ and $\sigma_2^2$, respectively. Moreover, the measurement equation is
\begin{equation}\begin{aligned}
z_{k\Delta t}=\begin{bmatrix}
{1} & {0} & {0} & {0}\\
{0} & {0.5} & {0} & {0}
\end{bmatrix}x_{k\Delta t}+v_{k}
\nonumber
\end{aligned}\end{equation}
where $\Delta t=0.02s$ is the sampling period, $v_k$ is the measurement noise with covariance $\mathrm{diag}(\sigma_3^2,\sigma_4^2)$. Moreover, the state is $x_{k\Delta t}=[x_1(k\Delta t);x_2(k\Delta t);\dot{x}_1(k\Delta t);\dot{x}_2(k\Delta t)]$. The system parameters are: $m_1=3kg$, $m_2=5kg$, $k_1=15N/m$, $k_2=5N/m$, $\sigma_1=\sigma_2=1$, $\sigma_3=0.5$ and $\sigma_4=0.1$.

The parameter matrix within each algorithm is set to $cI$, where $c$ is an adjustable scalar. The adjustable parameters of different methods are presented in Table I. Fig. 4 shows the performance metrics $E(300)$ and $T(300)$ versus the communication rate, respectively, from which it is clear that the proposed method has the best estimation performance at each of the communication rates. This further demonstrates the advantages of the proposed estimator and the corresponding posterior-based SET mechanism.

\begin{figure}[thpb]
      \centering
      \includegraphics[scale=0.7]{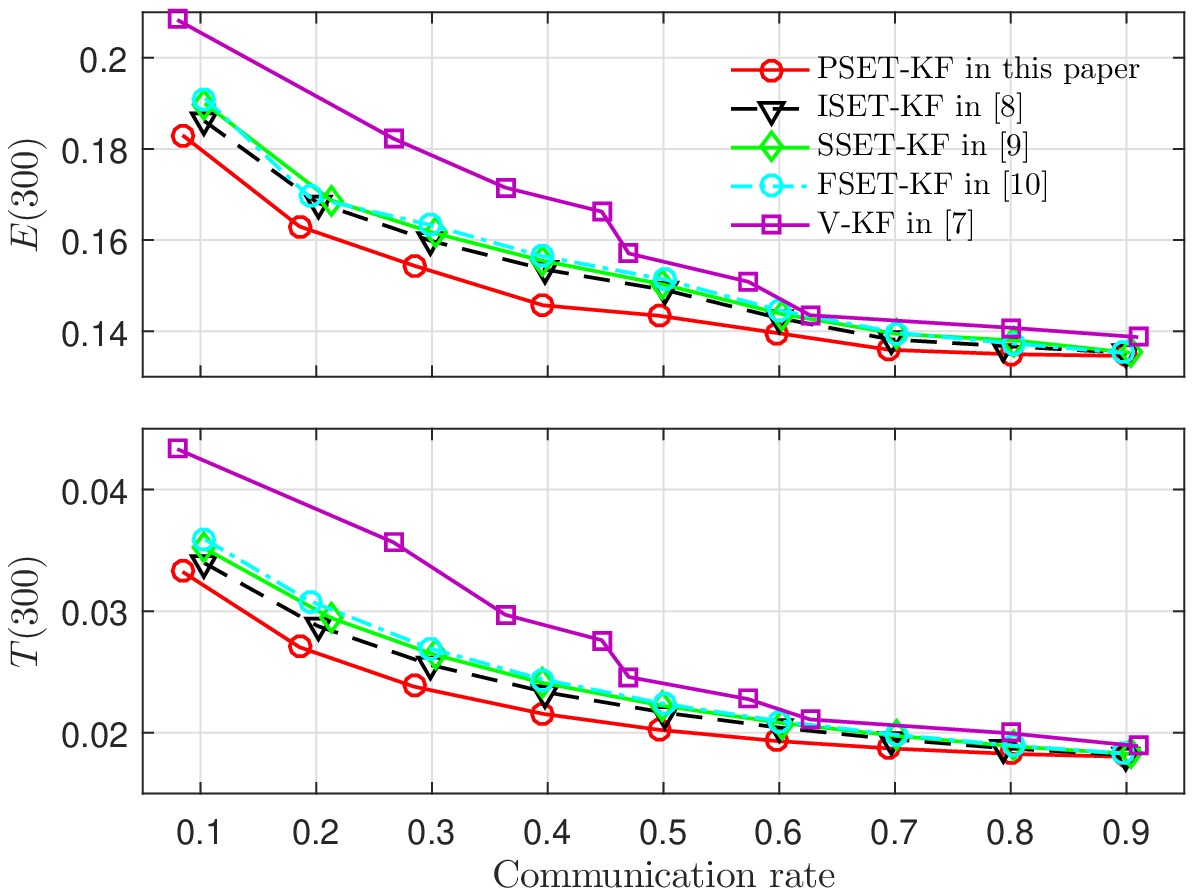}
      \caption{The estimation performances $E(300)$ and $T(300)$ of PSET-KF, ISET-KF, SSET-KF, FSET-KF and V-KF versus communication rate.}
\end{figure}
\section{Conclusion}
This paper proposed a posterior-based SET mechanism and derived the corresponding exact MMSE estimator. The proposed SET mechanism overcame the drawbacks of the innovation-based SET mechanism that ignored posterior and higher-order moment information, and thus had better data screening performance. The prediction error covariance of the designed estimator was proved to be asymptotically bounded. Moreover, based on the analytic performance results in the boundedness analysis, the expressions of the upper and lower bounds of the average communication rate were further derived, which guided selecting the key parameter matrix $\Gamma$ in the posterior-based SET mechanism. Finally, a target tracking system and a spring-mass system were provided to verify the advantages of the proposed methods.

In this paper, only the single-sensor case is considered, but for some large-scale systems, multiple sensors are often required to observe the whole state space. Therefore, our future work is to extend the proposed method to the multi-sensor case. In particular, how to analyze the priority of each sensor and the correlations among sensors will be the main challenges in this future work.


\appendix
\subsection{Proof of Lemma 1}
Adding $F$ and then subtracting $F$ leads to $E-E(E+F)^{-1}E=E(E+F)^{-1}(E+F-E)-F+F=(E(E+F)^{-1}-I)F+F=F+(E-E-F)(E+F)^{-1}F=F-F(E+F)^{-1}F=F(F^{-1}-(E+F)^{-1})F\geq 0$.
\subsection{Proof of Lemma 3}
It is well known that $\mathrm{Det}(XY)=\mathrm{Det}(YX)$ for any matrices $X$ and $Y$ with appropriate dimensions. Then, utilizing this property yields that $\mathrm{Det}(EF+I)=\mathrm{Det}(E^{1/2}FE^{1/2}+I)\geq \mathrm{Det}(E^{1/2}DE^{1/2}+I)=\mathrm{Det}(D^{1/2}ED^{1/2}+I)\geq \mathrm{Det}(D^{1/2}HD^{1/2}+I)=\mathrm{Det}(HD+I)$.
\subsection{The comparison between $\|\varepsilon_k\|^2_{K^T_k\Gamma K_k}$ and $\varrho_k$}
To analyze the effect of $\varrho_k$ on (31), we define $\gamma_k\triangleq\mathbb{E}[\|\varepsilon_k\|^2_{K^T_k\Gamma K_k}/\varrho_k]$. Obviously, a smaller $\gamma_k$ means that $\varrho_k$ plays a larger role, and a larger $\gamma_k$ means that $\varrho_k$ plays a smaller role. In other words, a larger $\gamma_k$ means that $\varrho_k$ has less influence on the trigger function (31) and the increased communication rate resulting from $\varrho_k$ is smaller.

\textbf{Lemma 4:} Consider the system (3) with SET mechanism (31), then one has
\begin{equation}\begin{aligned}
&\gamma_k=\mathbb{E}[\frac{\mathrm{Tr}(S P_{k|k-1}C^T(CP_{k|k-1}C^T+R)^{-1}CP_{k|k-1}S)}{\mathrm{Tr}(P^1_{k|k}\Gamma+P^0_{k|k}\Gamma-2(SP^0_{k|k}\Gamma P^1_{k|k}S)^{1/2})}].\nonumber
\end{aligned}\end{equation}

\textbf{Proof:} According to the law of iterated expectation, we have
\begin{equation}\begin{aligned}
\gamma_k=\mathbb{E}[\mathbb{E}[\frac{\|\varepsilon_k\|^2_{K^T_k\Gamma K_k}}{\varrho_k}|\mathcal{I}_{1:k-1}]].
\end{aligned}\end{equation}
The conditional expectation $\mathbb{E}[\frac{\|\varepsilon_k\|^2_{K^T_k\Gamma K_k}}{\varrho_k}|\mathcal{I}_{1:k-1}]$ in (60) can be simplified as
\begin{equation}\begin{aligned}
&\mathbb{E}[\frac{\|\varepsilon_k\|^2_{K^T_k\Gamma K_k}}{\varrho_k}|\mathcal{I}_{1:k-1}]\\
=&\frac{\mathrm{Tr}(\mathbb{E}[SK_k\varepsilon_k\varepsilon^T_kK^T_kS|\mathcal{I}_{1:k-1}])}{\varrho_k}\\
=&\frac{\mathrm{Tr}(SK_k\mathbb{E}[\varepsilon_k\varepsilon^T_k|\mathcal{I}_{1:k-1}]K^T_kS)}{\varrho_k}\\
=&\frac{\mathrm{Tr}(SP_{k|k-1}C^T(CP_{k|k-1}C^T+R)^{-1}CP_{k|k-1}S)}{\mathrm{Tr}(P^1_{k|k}\Gamma+P^0_{k|k}\Gamma-2(SP^0_{k|k}\Gamma P^1_{k|k}S)^{1/2})},
\end{aligned}\end{equation}
which completes the proof. $\square$

Recall the expressions of $P_{k|k-1}$, $P^0_{k|k}$ and $P^1_{k|k}$, one knows that (61) is a random variable determined by $\varsigma_1,\cdots,\varsigma_{k-1}$. In this case, according to Lemma 4 and the definition of mathematical expectation, one can easily deduce that
\begin{equation}\begin{aligned}
\gamma_k=&\mathbb{E}[\varphi(\Gamma,\varsigma_1,\cdots,\varsigma_{k-1})]\\
=&\sum_{\varsigma_1,\cdots,\varsigma_{k-1}} \varphi(\Gamma,\varsigma_1,\cdots,\varsigma_{k-1})\mathrm{Pr}(\varsigma_1,\cdots,\varsigma_{k-1}),
\end{aligned}\end{equation}
where
\begin{equation}\begin{aligned}
&\varphi(\Gamma,\varsigma_1,\cdots,\varsigma_{k-1})\\
\triangleq&\frac{\mathrm{Tr}(SP_{k|k-1}C^T(CP_{k|k-1}C^T+R)^{-1}CP_{k|k-1}S)}{\mathrm{Tr}(P^1_{k|k}\Gamma+P^0_{k|k}\Gamma-2(SP^0_{k|k}\Gamma P^1_{k|k}S)^{1/2})}.
\end{aligned}\end{equation}
Subsequently, there are two issues will be discussed. On one hand, notice that $\gamma_k$ is actually determined by $\Gamma$, thus the relationship between $\gamma_k$ and $\Gamma$ is of interest to us. On the other hand, the relationship between $\gamma_k$ and 1 is also a matter of interest, since it reflects whether $\|\varepsilon_k\|^2_{K^T_k\Gamma K_k}$ or $\varrho_k$ plays a major role in (31). Unfortunately, the above problems are difficult to be solved perfectly due to the closed-form expression of $\gamma_k$ is not available. Therefore, we can only explore the above issues through some heuristic theoretical derivations and Monte Carlo simulation. The latter analysis is roughly divided into the following three steps:
\begin{itemize}
\item First, we discuss two special extreme cases theoretically, i.e., $\Gamma$ is very small (tends to be $0$) and $\Gamma$ is very large (tends to be $\infty$).
\item Then, we explore the relationship between $\Gamma$ and $\gamma_k$ by Monte Carlo simulation. Meanwhile, the relationship between $1$ and $\gamma_k$ will also be explored by Monte Carlo simulation.
\item Finally, based on the theoretical derivation and Monte Carlo simulation, we summarize some of the properties that $\gamma_k$ may possess.
\end{itemize}

Consider a special case with $\Gamma=aI$, then (63) can be simplified as
\begin{equation}\begin{aligned}
&\varphi(aI,\varsigma_1,\cdots,\varsigma_{k-1})\\
=&\frac{\mathrm{Tr}(P_{k|k-1}C^T(CP_{k|k-1}C^T+R)^{-1}CP_{k|k-1})}{\mathrm{Tr}(P^1_{k|k}+P^0_{k|k}-2(P^0_{k|k}P^1_{k|k})^{1/2})}.
\end{aligned}\end{equation}

On one hand, when $a\rightarrow\infty$, $\Gamma$ is very large and the communication rate tends to be $1$. In this case, $\mathrm{Pr}(\varsigma_1=1,\cdots,\varsigma_{k-1}=1)\rightarrow1$, and $\gamma_k$ can be approximated as
\begin{equation}\begin{aligned}
\gamma_k\approx&\lim_{a\rightarrow\infty}\varphi(aI,\varsigma_1=1,\cdots,\varsigma_{k-1}=1).
\end{aligned}\end{equation}
According to the expressions of $P^0_{k|k}$ and $P^1_{k|k}$, one knows that $P^0_{k|k}\rightarrow P^1_{k|k}$ when $\Gamma\rightarrow \infty I$. This means
\begin{equation}\begin{aligned}
\mathrm{Tr}(P^1_{k|k}+P^0_{k|k}-2(P^0_{k|k}P^1_{k|k})^{1/2})\rightarrow0.
\end{aligned}\end{equation}
Meanwhile, notice that $P_{k|k-1}=AP_{k-1|k-1}A^T+Q\geq Q>0$, which means
\begin{equation}\begin{aligned}
&\mathrm{Tr}(P_{k|k-1}C^T(CP_{k|k-1}C^T+R)^{-1}CP_{k|k-1})\\
\geq&\frac{\mathrm{Tr}(CP_{k|k-1}C^T(CP_{k|k-1}C^T+R)^{-1}CP_{k|k-1}C^T)}{\lambda_{max}(C^TC)}\\
\geq &\frac{\mathrm{Tr}(CP_{k|k-1}C^T(CP_{k|k-1}C^T+\lambda_{max}(R)I)^{-1}CP_{k|k-1}C^T)}{\lambda_{max}(C^TC)}\\
\geq &\frac{\mathrm{Tr}(CQC^T(CQC^T+\lambda_{max}(R)I)^{-1}CQC^T)}{\lambda_{max}(C^TC)}>0,
\end{aligned}\end{equation}
where the last inequality follows from Lemma 5 in Appendix D. Then, it follows from (66) and (67) that
\begin{equation}\begin{aligned}
\lim_{a\rightarrow\infty}\varphi(aI,\varsigma_1=1,\cdots,\varsigma_{k-1}=1)\rightarrow\infty.
\end{aligned}\end{equation}
This means $\gamma_k$ is likely to be very large and $\varrho_k$ plays almost no role (very small role) in the trigger function (31).

On the other hand, when $a\rightarrow0$, $\Gamma$ is very small and the communication rate tends to be $0$. In this case, $\mathrm{Pr}(\varsigma_1=0,\cdots,\varsigma_{k-1}=0)\rightarrow1$ and $\gamma_k$ can be approximated as
\begin{equation}\begin{aligned}
\gamma_k\approx&&\lim_{a\rightarrow0}\varphi(aI,\varsigma_1=0,\cdots,\varsigma_{k-1}=0).
\end{aligned}\end{equation}
Meanwhile, notice that,
\begin{equation}\begin{aligned}
&\lim_{a\rightarrow0}\varphi(aI,\varsigma_1=0,\cdots,\varsigma_{k-1}=0)\\
=&\frac{\mathrm{Tr}(P_{k|k-1}C^T(CP_{k|k-1}C^T+R)^{-1}CP_{k|k-1})}{\mathrm{Tr}(P^1_{k|k}+P_{k|k-1}-2(P_{k|k-1}P^1_{k|k})^{1/2})}.
\end{aligned}\end{equation}
Then, two cases will be discussed. When $\rho(A)<1$, $P_{k|k-1}$ must be bounded. In this case, according to Corollary 3 in \cite{WD Gaussian}, one knows $\mathrm{Tr}(P^1_{k|k}+P_{k|k-1}-2(P_{k|k-1}P^1_{k|k})^{1/2})>0$ and
\begin{equation}\begin{aligned}
0<\lim_{a\rightarrow0}\varphi(aI,\varsigma_1=0,\cdots,\varsigma_{k-1}=0)<\infty,
\end{aligned}\end{equation}
i.e., $\lim_{a\rightarrow0}\varphi(aI,\varsigma_1=0,\cdots,\varsigma_{k-1}=0)$ must be a positive real number. When $\rho(A)\geq 1$ and $\Gamma\rightarrow0$, $P_{k|k-1}$ may tend to infinite. In this case, it is difficult to analyze whether (71) holds. Fortunately, when both $A$ and $C$ are scalars, the above problem can be solved. Therefore, we decide to analyze a scalar system. When $A$ and $C$ are scalars, by taking a bit of tedious but straightforward simplifications, one has
\begin{equation}\begin{aligned}
&\lim_{a\rightarrow0}\varphi(aI,\varsigma_1=0,\cdots,\varsigma_{k-1}=0)\\
=&\frac{\frac{C^2P^2_{k|k-1}}{C^2P_{k|k-1}+R}}{-\frac{C^2P^2_{k|k-1}}{C^2P_{k|k-1}+R}+2P_{k|k-1}-2(P^2_{k|k-1}-\frac{C^2P^3_{k|k-1}}{C^2P_{k|k-1}+R})^{1/2}}\\
=&\frac{C^2}{C^2+2\frac{R}{P_{k|k-1}}-2((C^2+\frac{R}{P_{k|k-1}})\frac{R}{P_{k|k-1}})^{1/2}}.
\end{aligned}\end{equation}
Notice that $P_{k|k-1}>Q$, thus we only need to discuss the case $P_{k|k-1}\rightarrow\infty$ (since there is no singularity
in (72) within $(Q,\infty)$, and if (72) is a real number when $P_{k|k-1}\rightarrow\infty$, then (72) will also be a real number when $P_{k|k-1}<\infty$). Then, let $P_{k|k-1}\rightarrow\infty$, one can derive
\begin{equation}\begin{aligned}
0<\lim_{P_{k|k-1}\rightarrow\infty,a\rightarrow0}\varphi(aI,\varsigma_1=0,\cdots,\varsigma_{k-1}=0)=1<\infty
\end{aligned}\end{equation}
from (72). This means $\gamma_k$ is likely to be a positive real number, and both $\|\varepsilon_k\|^2_{K^T_k\Gamma K_k}$ and $\varrho_k$ play a role in the trigger function (31).

When $\Gamma$ does not belong to the above two special extreme cases, it is difficult to analyze $\gamma_k$ theoretically (since the closed-form of $\gamma_k$ is not available). To remedy this part, Monte Carlo simulation will be provided. Consider the target tracking system used in Section IV-A, then we approximate $\gamma_k$ by 500 Monte Carlo trails. Particularly, all parameters are consistent with Section IV-A. Fig. 5 shows the relationship between $\gamma_k$ and the communication rate. Then, from Fig. 5 and Table I we can see that $\gamma_k$ increases as the adjustable parameter $\Gamma$ increases. This means that the larger the adjustable parameter $\Gamma$, the smaller the role played by $\varrho_k$ in the trigger function (31). In fact, the simulation result is also consistent with our previous theoretical analysis, i.e., $\varrho_k$ plays almost no role when $\Gamma$ tends to infinity, while $\varrho_k$ plays a role when $\Gamma\rightarrow0$.
\begin{figure}[thpb]
      \centering
      \includegraphics[scale=0.7]{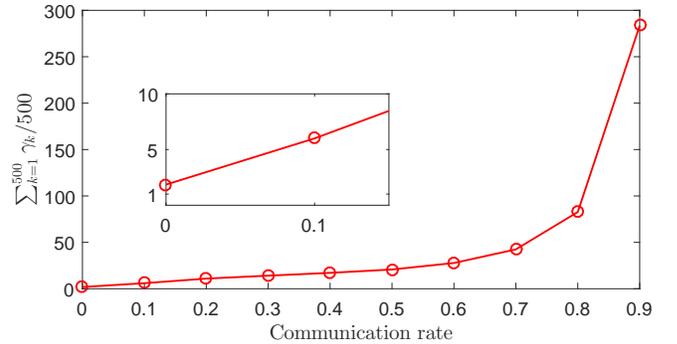}
      \caption{The relationship between $\gamma_k$ and communication rate.}
\end{figure}

Based on the above theoretical derivation and Monte Carlo simulation, we summarize the following phenomena:
\begin{itemize}
\item A larger $\Gamma$ (or communication rate) tends to imply that $\varrho_k$ plays a smaller role and $\|\varepsilon_k\|^2_{K^T_k\Gamma K_k}$ plays a larger role in the event-triggered mechanism (31).
\item If $\Gamma$ is very large ($\Gamma\rightarrow\infty$, communication rate $\rightarrow1$), then $\gamma_k$ tends to be infinite, and only $\|\varepsilon_k\|^2_{K^T_k\Gamma K_k}$ is working, while $\varrho_k$ hardly works. If $\Gamma$ is very small ($\Gamma\rightarrow0$, communication rate $\rightarrow0$), then $\gamma_k$ tends to be some positive real numbers, and both $\|\varepsilon_k\|^2_{K^T_k\Gamma K_k}$ and $\varrho_k$ play a role in the event-triggered mechanism (31).
\end{itemize}

Next, we discuss the relationship between $\gamma_k$ and $1$. Similar to the above analysis, we analyze here a special case where $A$ and $C$ are scalars and $\Gamma=aI$ ($a\rightarrow0$). Then, according to (72) we can easily obtain
\begin{equation}\begin{aligned}
&\lim_{a\rightarrow0}\varphi(aI,\varsigma_1=0,\cdots,\varsigma_{k-1}=0)\\
=&\frac{C^2}{C^2+2\frac{R}{P_{k|k-1}}-2((C^2+\frac{R}{P_{k|k-1}})\frac{R}{P_{k|k-1}})^{1/2}}\\
\geq& \frac{C^2}{C^2+2\frac{R}{P_{k|k-1}}-2(\frac{R}{P_{k|k-1}}\frac{R}{P_{k|k-1}})^{1/2}}=1.
\end{aligned}\end{equation}
Then, by combining the above phenomena that $\gamma_k$ increases with $\Gamma$, we reasonably suspect that $\gamma_k$ tends to be larger than $1$. In fact, we can also see from Fig. 5 that $\gamma_k$ tends to be larger than $1$. Therefore, $\varrho_k$ is an auxiliary role in the trigger function (31), and it is $\|\varepsilon_k\|^2_{K^T_k\Gamma K_k}$ that really dominates.

\textbf{Remark 3:} It is necessary to emphasize that this subsection only makes some conjectures about the properties of $\gamma_k$ through Monte Carlo simulation and some heuristic theoretical derivations, rather than giving deterministic theorems.
\subsection{Proof of Lemma 5}
\textbf{Lemma 5:} If $a\in\mathbb{R}^1$, $a>0$, $X,Y\in\mathbb{R}^{n\times n}$ and $Y\geq X\geq 0$, then $\mathrm{Tr}(X(X+aI)^{-1}X)\leq\mathrm{Tr}(Y(Y+aI)^{-1}Y)$.

\textbf{Proof:} Denote the singular value decomposition of $X$ and $Y$ as
\begin{equation}\begin{aligned}
X=S_XV_XS_X^T,\ Y=S_YV_YS^T_Y.
\end{aligned}\end{equation}
Then, it can be easily deduce that
\begin{equation}\begin{aligned}
&X(X+aI)^{-1}X=S_XV_X(V_X+aI)^{-1}V_XS_X^T\\
&Y(Y+aI)^{-1}Y=S_YV_Y(V_Y+aI)^{-1}V_YS_Y^T.
\end{aligned}\end{equation}
According to (78), one has
\begin{equation}\begin{aligned}
&\mathrm{Tr}(X(X+aI)^{-1}X)=\mathrm{Tr}(V_X(V_X+aI)^{-1}V_X)\\
&\mathrm{Tr}(Y(Y+aI)^{-1}Y)=\mathrm{Tr}(V_Y(V_Y+aI)^{-1}V_Y).
\end{aligned}\end{equation}
Notice that, $V_X$ and $V_Y$ are diagonal matrices. In this case, using Theorems 10.21 and 10.22 in \cite{Matrix theory} yields $V_Y\geq V_X$. Finally, it follows from the fact $f(x)=\frac{x^2}{x+1}$ is monotonically increasing on $[0,+\infty)$ that the lemma can be obtained. $\square$
\ifCLASSOPTIONcaptionsoff
  \newpage
\fi



\end{document}